\documentclass[twocolumn,prb,amsmath,showpacs,amssymb,citeautoscript,floatfix]{revtex4}
\usepackage{graphicx}

\usepackage{color}
                
\def\etal{{ \it et al. }}
\def\prb{{Phys. Rev. B }}

\usepackage{dcolumn}
\usepackage{bm}
\usepackage{amsmath}
\begin{document}

\title{Pressure-dependence of Curie temperature and resistivity in 
complex Heusler alloys} 

\date{\today}

\author{S.K. Bose}
\affiliation{Department of Physics, Brock University,
St. Catharines, Ontario, Canada, L2S 3A1}

\author{J. Kudrnovsk\'y}
\affiliation{Institute of Physics, Academy of Sciences of the
Czech Republic, CZ-182 21 Praha 8, Czech Republic}

\author{V. Drchal} 
\affiliation{Institute of Physics, Academy of Sciences of the
Czech Republic, CZ-182 21 Praha 8, Czech Republic}

\author{I. Turek}
\affiliation{Charles University, Faculty of Mathematics and 
Physics, Department of Condensed Matter Physics, Ke Karlovu 5, 
CZ-12116 Prague 2, Czech Republic}

\date{\today}

\begin{abstract}
Using first-principles electronic structure calculations,
we have studied the dependence of the Curie temperature on external
hydrostatic pressure for random Ni$_{2}$MnSn Heusler alloys doped
with Cu- and Pd-atoms, over the entire range of dopant concentrations.
The Curie temperatures are calculated by applying 
random-phase approximation to the Heisenberg Hamiltonian whose
parameters are determined using the linear response and multiple 
scattering methods, based on density-functional theory.
In (Ni$_{1-x}$,Pd$_x$)$_{2}$MnSn alloys the Curie temperature is found to increase 
with applied pressure over the whole concentration range.
The crossover from the  increase to  the decrease of the Curie 
temperature with pressure takes place for Cu-concentrations larger than
about 70\% in (Ni$_{1-x}$,Cu$_x$)$_{2}$MnSn Heusler alloys.
The results for the reference Ni$_{2}$MnSn Heusler alloy agree well
with a previous theoretical study of Sasioglu {\it et al.} and
also reasonably well with available experimental data.
Results for the spin-disorder induced part of the resistivity
in (Ni$_{1-x}$,Pd$_x$)$_{2}$MnSn Heusler alloys, calculated by  using the 
disordered local moment model, are also presented.
 Finally, a qualitative understanding of the results, based on 
Anderson's superexchange interaction and Stearn's model of the indirect exchange interaction between
localized and itinerant $d$-electrons, is provided.

\end{abstract}


\maketitle

\section{Introduction}
High pressure studies form an important area of research in solid state
physics. 
Over the years such studies have provided  useful insight into our
understanding 
of the physical properties of solids\cite{p-rev,sc-nature}. 
Pressure has basically two  effects on the electrons in solids: increased kinetic
energy and accompanying
change in the effects of Coulomb interaction among the electrons. In typical
band structure
calculations the former effect is captured via increased overlap of the basis
orbitals, leading to increased
hopping matrix elements and band broadening. The changes in the correlation
effects are captured via
self-consistency of charge and potential, as dictated by the density functional
theory and its variants
(e.g. generalized gradient approximation (GGA), local density approximation with
on-site Coulomb interaction (LDA+U)).
Both of these affect magnetic properties of materials such as exchange interaction,
local moments, Curie temperature and resistivity due to magnetic scattering. 
Of particular importance for metallic magnets are the change in the Fermi surface and
the hybridization between different orbitals, which influences the itinerant vs. 
localized nature of the charge 
carriers. It is a useful exercise to explore how the existing theories of
electronic and magnetic structure calculation fare in 
describing the pressure variation of these physical properties. 
In the present paper we  study
 the pressure-dependence of the Curie temperature and the resistivity of some  
quaternary Heusler alloys, which form an important 
class of magnetic materials, with potential industrial/technological 
applications\cite{review,sme,gmce}. 
The present  work is a natural continuation of our previous paper \cite{qha_th}
in which we presented an extensive study of the magnetic and transport properties 
of these alloys at ambient pressure.

There exists a number of experimental studies of the variation 
of the Curie temperature (T$_{c}$) under pressure
for elemental ferromagnets, transition-metal alloys
\cite{tc_nimn,tc_tmall}, as well as Heusler alloys \cite{tc_pr3,tc_pr4,tc_pr5}
including Co$_{2}$TiAl, \cite{tc_pr4} Ni$_{2}$MnSn, \cite{tc_pr3,tc_pr5}
Pd$_{2}$MnSn and some others \cite{tc_pr3}.  Recently 
diluted magnetic semiconductors \cite{tc_dms} and some random Heusler
alloys \cite{rha_pr} have also been studied.
However, the corresponding experimental studies of the
pressure-dependence of the resistivity are rare: one example of such a study 
is the work by Austin and Mishra \cite{tc_pr3}.

Several studies of the ambient/equilibrium electronic and magnetic properties of such alloys have appeared
recently\cite{tc_niall,tc_gamnas,ha-sha,qsha1,qsha2,qha_th}.
Theoretical studies of the pressure-dependence of the Curie 
temperature of elemental transition metal ferromagnets include model-based studies 
\cite{tc_pr1,tc_pr2} as well as recent first-principles calculations.
\cite{tc_fe1,tc_fe2} 
Turek \etal\cite{Turek2003} have studied the pressure-dependence 
of the Curie temperature in hcp Gd and found a strong dependence of 
$T_{c}$ on the $c/a$ ratio, which, they suggested, can be related to 
the measured $T_{c}$ of thick epitaxial Gd(0001) films on various 
transition-metal substrates.
Similar calculations have also been carried out for the intermetallic
compound GdAl$_2$ \cite{gdal2} and for Cr-based compounds in the
zinc blende structure. \cite{crx}
A systematic theoretical study of the 
pressure-dependence of $T_{c}$ in Heusler alloys was carried out recently by
Sasioglu {\it et al.}, \cite{tc_ni2mnsn} where the experimentally
observed increase of $T_{c}$ with pressure \cite{tc_pr3,tc_pr5} in
Ni$_{2}$MnSn was reproduced correctly, albeit only qualitatively so.
There exists some other studies of Heusler 
alloys in which the dependence of the exchange integrals on volume
contraction/expansion \cite{antrop,meinert,sasioglu,qsha2} was explored.

The particular Heusler systems considered in this work are the (Ni$_{1-x}$,Cu$_x$)$_{2}$MnSn
and (Ni$_{1-x}$,Pd$_x$)$_{2}$MnSn alloys.
We employ the coherent potential approximation (CPA) to treat the disorder in the Ni-sublattice 
due to doping with Cu- or Pd-atoms. 
 It is well known that  CPA neglects the effects of short-range order, which
are often present in random alloys.
However, for the  cases under study, such effects should be weak, as  we need 
to describe  the magnetic interactions on the non-random Mn sublattice, which is 
only indirectly influenced by the randomness of the transition-metal sublattice 
via hybridization. The most important effect of disorder is, therefore, 
the change in carrier  concentration, which is  well-described by the CPA.
In a recent paper  \cite{cpa_sc}, it was demonstrated that the CPA provides
a reasonably good description of the electronic structure for the case
of disorder on the magnetic sublattice as well.

\section{Formalism}
\label{Form}

The electronic structure calculations are performed employing the 
tight-binding linear muffin-tin orbital (TB-LMTO) basis \cite{lmto} 
and the density functional theory (DFT).  The local spin-density
approximation (LSDA) for the exchange-correlation part of the potential
is used. 
The effect of substitutional disorder among (Ni,Cu) or (Ni,Pd) 
atoms is described by the coherent potential approximation (CPA)
as formulated in the framework  of the TB-LMTO Green function  
method. \cite{book}
The calculations employ  an $s,p,d,f$-basis, the same atomic radii
are adopted for all atoms, and the Vosko-Wilk-Nusair exchange-correlation 
potential \cite{VWN} is used.

The exchange interactions and Curie temperatures are
studied here by employing a classical Heisenberg Hamiltonian
\begin{equation}
H = - {\sum_{i \neq j}} J_{ij} \, {\bf e}_i \cdot {\bf e}_j ,
\label{eqhh}
\end{equation}
where  $J_{ij}$ denotes the exchange integral between
Mn atoms at sites $i$ and $j$, and  ${\bf e}_i$ and
${\bf e}_j$ are unit vectors in the directions of the local
magnetization on sites $i$ and $j$, respectively.   
The parameters of the Heisenberg Hamiltonian are obtained
using a two-step approach. \cite{lie,eirev} 
The reference state for our calculations is chosen to be the
disordered local moment (DLM) state, \cite{dlm} which was successfully
used in previous studies. \cite{qsha2,qha_th}  
The exchange interactions $J_{ij}$ can be expressed in the framework
of the TB-LMTO method as
\begin{equation}
J_{ij} = \frac{1}{4\pi} \, \mathrm{Im} \, \int_{C}  \mathrm {tr}_{L}
\{ \Delta_{i}(z) \, \bar{g}^{\uparrow}_{ij}(z) \,
\Delta_{j}(z) \, \bar{g}^{\downarrow}_{ji}(z) \}
\, \mathrm{d}z ,
\label{eqjij}
\end{equation}
where $\Delta_{i}(z)$ characterizes the exchange splitting of
the Mn atom at the site $i$, $\bar{g}^{\sigma}_{ij}(z)$  is
the configurationally averaged Green function describing the motion of
an electron with spin $\sigma$, $\sigma=\uparrow, \downarrow$,
between Mn sites $i$ and $j$, and the integration is done over
the contour $C$ in the complex energy plane $z$ which starts below the
valence band and ends at the Fermi energy.
Symbol $\mathrm{tr}_L$ denotes the trace over the basis orbitals of angular momentum symmetry
$L\equiv (\ell,m)$.
The effective exchange splitting $\Delta_{i}(z)$ is defined in the
TB-LMTO method in terms of the potential functions
$P_{i}^{\,\sigma}(z)$ of Mn atoms as
$\Delta_{i}(z)=P^{ \uparrow}_{i}(z) - P^{ \downarrow}_{i}(z)$.
Further details on the exchange interactions evaluated in the DLM
reference state can be found in Ref.~\onlinecite{eidlm}.

We have found that the mapping can be improved (providing a better agreement 
of calculated $T_{c}$ with the experiment) by including the 
electron correlation effect on Ni-sites in the framework of the 
LSDA+U method (see Fig.~1 in Ref.~\onlinecite{qha_th}).
The electron correlations on Pd-sites are found to have negligible influence on the calculated $T_{c}$.
We determine the Curie temperature corresponding to the Heisenberg 
Hamiltonian using the RPA \cite{eirev} and by including all
exchange interactions up to $\sim$ 4 lattice constants.
The pressure is simulated by reduction of the lattice constants
from their ambient pressure values, as reported in experiments 
(see Table 4 of Ref.~\onlinecite{qha1} and Table 2
of Ref.~\onlinecite{qha2}).
In Ref.~\onlinecite{tc_pr5} an empirical pressure-volume relation  
was given for Ni$_{2}$MnSn. 
Although such a relation can be obtained entirely from theoretical
calculations, 
in the present study, as in Ref.~\onlinecite{tc_ni2mnsn}, we simply 
reduce the lattice constant and estimate the corresponding pressure 
from the empirical pressure-volume relation. 
According to both the empirical pressure-volume relation\cite{tc_pr5} and theoretical
calculations\cite{tc_ni2mnsn}, a 3~\% reduction of the lattice constant 
corresponds  roughly  to a pressure of 16~GPa in Ni$_{2}$MnSn.

In Heusler alloys (Ni$_{1-x}$,T$_x$)$_{2}$MnSn (T=Cu, Pd) the effect of
disorder among the Ni- and T-atoms at the Fermi energy $E_F$ is 
weak  because the Ni- and T-states lie well below $E_F$. \cite{qha_th}
Consequently, the corresponding residual resistivity is small. 
The resistivity due to phonon scattering \cite{resT} is known to be 
small as well.
The dominant contribution to the resistivity, thus, is from the 
spin-disorder scattering, which increases with temperature, reaching 
its maximum value at $T_{c}$.
In Ref.~\onlinecite{qha_th} we employed a simplified approach 
to estimate the temperature-dependent resistivity in 
(Ni$_{1-x}$,Pd$_x$)$_{2}$MnSn Heusler alloys. 
It was shown  that the spin-disorder at and above $T_{c}$ can be 
described satisfactorily using the DLM model, and the corresponding 
resistivity can be described by the Kubo-Greenwood approach 
adapted to the TB-LMTO method. \cite{kglmto}
We have used the same approach to study the resistivity under 
pressure as well. For further details concerning computational techniques  
we refer the reader to our recent paper. \cite{qha_th}

The method we have used is based on mapping the ($T=0$) total energy to an effective Heisenberg Hamiltonian.
In fact it is just the band energy, not the total energy, which is considered in the mapping, by appealing to
the so-called 'magnetic force theorem'(see Ref.(\onlinecite{crx}) and references therein).
For finite temperature ($T\neq 0$) calculations, as would be appropriate for the DLM reference state, one needs to
consider the mapping of the free energy $F=U-TS$, where $U$ is the internal energy and $S$ is the entropy.
$U$ should include the electronic and the average vibrational energy and $S$ should include both electronic and
vibrational entropy. Electronic energy should be calculated via finite temperature density functional theory.
This free energy cannot, in principle, be mapped to an effective Heisenberg form, which includes only the electronic 
parameter (spin or magnetic moment). Such a scheme could be implemented for a 'supercell' calculation. However,
the advantage of being able to calculate exchange interactions up to a large distance with relative ease and
reasonable accuracy and therefore predict results and trends for a group of materials will be gone. For finite temperatures,
both electronic excitations and lattice vibrations will influence magnetic properties. However, in the absence of an
accurate scheme, we will refrain from speculating on these effects on our results.

\section{Results and discussion}

In this section we present results for the pressure-dependence of the Curie temperature
and spin-disorder  resistivity 
of (Ni$_{1-x}$,T$_x$)$_{2}$MnSn (T=Cu, Pd) Heusler alloys over a broad range 
of concentrations.

\subsection{Curie temperature under pressure: previous studies}
\label{CP}
We start with a brief review of previous theoretical studies, both model-based 
and first-principles. Theoretical study of the pressure-dependence of $T_{c}$ on a model
level has usually been  based on a discussion of the indirect exchange 
interaction among spins \cite{tc_pr1} or on the Hubbard model of 
magnetism. \cite{tc_pr2}
It is clear that the pressure-dependence of $T_{c}$ carries
information about the volume-dependence of the electron wave functions 
\cite{tc_pr1} (Bloch functions in the case of crystalline solids). 
In an RKKY-type indirect exchange interaction model, the volume-dependence 
of $T_{c}$ can be related to the volume dependencies of the
density of states at the Fermi level $N(E_F)$ and the exchange
integral or the matrix element of the exchange interaction between 
the wave functions at the Fermi energy \cite{tc_pr1}(henceforth referred 
to as the bare exchange interaction $J^{\rm bare}$, see Eq.(\ref{eqjbare})). 
The density of states at the Fermi level usually decreases with 
pressure (unless, under pressure, new bands start crossing the Fermi 
level), as the bands broaden due to increased overlap between the 
orbitals centered on neighboring atoms. 
The bare exchange integral usually increases under pressure, due to 
increased overlap of the wave functions. The above two effects compete with each other
in determining the final change of $T_c$ under pressure.

In a description based on the Hubbard model, the band terms and the 
term involving the intra-atomic Coulomb interaction $U$ are written
separately. 
In this case it is possible to divide the volume/pressure-dependence 
of $T_{c}$ into a term originating from the band broadening under 
pressure and another that relates to the volume-dependence of $U$. 
The importance of this latter term for the pressure-dependence of
$T_{c}$ in Ni and Ni-Cu alloys was first pointed out by Lang and 
Ehrenreich. \cite{tc_pr2}
They argued that it is this term involving the intra-atomic Coulomb 
interaction that is responsible for the increase in $T_{c}$ under 
pressure ($P$) in Ni and the decrease of $dT_c/dP$ with the increase 
of Cu-concentration in Ni-Cu alloys.

In {\it ab initio} calculations based on DFT, such as ours,  
the above effects are merged together and it is hard to delineate one 
from the other. 
The advantage, of course, is that the exchange interactions are 
realistically calculated for the system at hand, and no assumption 
related to the volume-dependence of any parameter needs to be made. 
Prototypical studies in this direction are Refs.~\onlinecite{tc_fe1,tc_fe2},
addressing the pressure-dependence of the Curie temperature of bcc iron. 
In particular, a very recent first-principles study \cite{tc_fe2} 
seems to agree with experiment, which gives essentially no pressure-dependence
 of $T_{c}$.
The authors describe this result as a balance between the band structure 
effect which reduces the magnetization and an increase of the bare exchange 
integrals with decreasing volume, as mentioned above.
A somewhat different situation is encountered for fcc-Ni. 
This metal is a strong ferromagnet, its magnetic moment depends on 
pressure only weakly and an increase of exchange integrals under 
pressure thus dominates. As a result, $T_c$ of fcc-Ni increases with pressure.
In general both effects are relevant and the observed change of $T_c$ under pressure is
a result of the competition between them.

\begin{figure}
\center \includegraphics[width=8.0 cm]{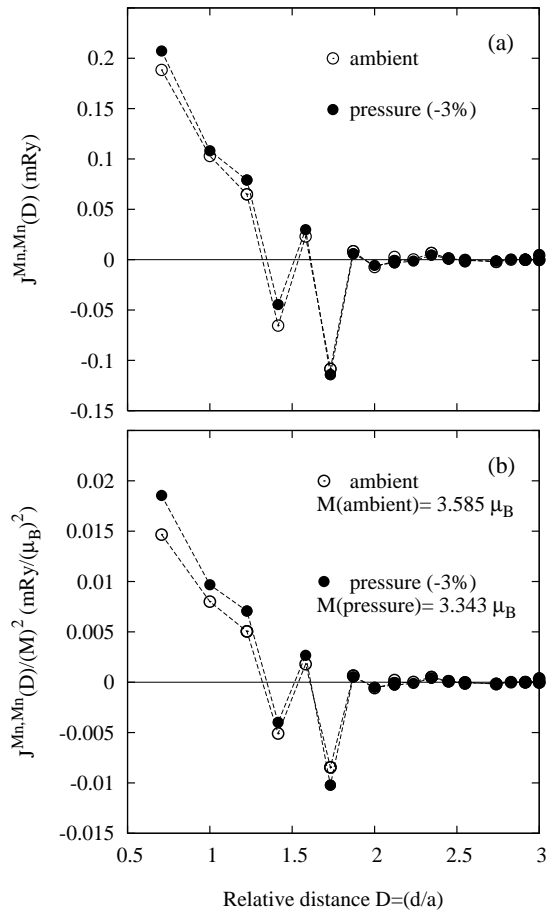}
\caption {
The Mn-Mn exchange interactions in the Ni$_{2}$MnSn Heusler alloy as
functions of the relative interatomic distance $d/a$ where $a$ is the
fcc lattice parameter:
(a) the effective interactions $J_{ij}$ calculated from
Eq.~(\ref{eqjij}), 
(b) the bare interactions $J_{ij}^{\rm bare}$ defined by
Eq.~(\ref{eqjbare}).
The cases of ambient pressure (open circles) and the pressure
corresponding to the 3\% reduction of the lattice constant
(filled circles) are shown. 
}
\label{fX}
\end{figure}

The above-mentioned competition of the two effects is illustrated 
in Fig.~\ref{fX}.
Here we plot the effective
exchange interactions $J_{ij}$ ( Eq.\ref{eqjij}), which
control directly the Curie temperature of the system (Fig.~\ref{fX}a),
and the bare exchange interactions $J_{ij}^{\rm bare}$, defined as
\begin{equation} 
J_{ij}^{\rm bare} = J_{ij}/(M_i M_j) ,
\label{eqjbare}
\end{equation}
 as a function 
of the distance between the Mn atoms in
Ni$_{2}$MnSn.  We consider the ambient pressure case, as well as elevated 
pressure simulated by 
reduction of the lattice constant. 
In Eq.(\ref{eqjbare}) $M_i$ denotes the size of the local magnetic moment of Mn atom
at site $i$ (Fig.~\ref{fX}b).
This definition is motivated by the presence of the exchange
splittings $\Delta_i(z)$ in Eq.~(\ref{eqjij}) which are roughly
proportional to the moment magnitudes $M_i$.
The increase of the bare exchange interactions with
pressure is clearly seen for the first five neighbors (see Fig.~\ref{fX}b).
It is also obvious that magnetic moments decrease with the pressure,
as expected.
These two effects, i.e., enhancement of  $J_{ij}^{\rm bare}$ 
and suppression of magnetic moments, both due to the volume  decrease, 
compete  with each other, resulting in smaller pressure-induced
changes of the effective interactions $J_{ij}$ (Fig.~\ref{fX}a).   
The final result in the present case is an enhancement of $T_{c}$
under hydrostatic pressure.
The same qualitative explanation was put forward by Csontos \etal \cite{tc_dms} for the 
increase of $T_{c}$ with pressure in some diluted magnetic 
semiconductors.

\subsection{Calculated results}

In this section we present and compare results for (Ni, Pd)$_{2}$MnSn and 
(Ni, Cu)$_{2}$MnSn Heusler alloys, with
Ni$_{2}$MnSn  considered as the  reference case.

\subsubsection{Ni$_{2}$MnSn and Pd$_{2}$MnSn Heusler alloys}

The Heusler alloy Ni$_{2}$MnSn was studied theoretically in great 
detail in a recent paper, \cite{tc_ni2mnsn} while corresponding 
experimental results can be found in Refs.~\onlinecite{tc_pr3}  
and ~\onlinecite{tc_pr5}.
The experiments give the rate of increase of $T_{c}$ with pressure 
around 0.62~K/kbar \cite{tc_pr3} (for pressures up to 10~kbar or 1~GPa) 
and 0.744~K/kbar for higher pressures up to 9~GPa. \cite{tc_pr5}
The authors of the theoretical study \cite{tc_ni2mnsn} employ the 
frozen-magnon approach and inverse lattice Fourier  transform to 
construct the effective Heisenberg Hamiltonian, which is then analyzed 
in the framework of the multisublattice MFA.
They obtain an increase of $T_{c}$ by 38~K for pressures around 16~GPa,
an increase from the ambient value $T_{c}$=362~K to $T_{c}$=400~K.
Our estimated values are $T_{c}$=322 (334)~K for ambient pressure and 
$T_{c}$=375 (401)~K (an increase of 53 (67)~K for 3\% reduction of 
the lattice constant, roughly corresponding to the same pressure 
$\sim$ 16~GPa).
The values in brackets correspond to the model assuming electron
correlations on Ni-sites (the effective Hubbard $U_{\rm Ni}$  chosen is 2~eV).
Assuming linear relation between $T_{c}$ and pressure, the experimental
increase is expected to be around 100$-$120~K.
The theory thus correctly predicts a pressure-induced increase of
$T_{c}$, while the absolute increase is at least three times 
(Ref.\onlinecite{tc_ni2mnsn}) to 1.5$-$1.8-times (present study) too small.
The experimental data for Pd$_{2}$MnSn Heusler alloys  that we have 
come across in the literature \cite{tc_pr3} also indicate an increase 
of $T_{c}$ 
with pressure at a rate of 0.75~K/kbar, the same \cite{tc_pr5}
or almost the same \cite{tc_pr3} as that for Ni$_{2}$MnSn.
We have  correctly obtained an increase of $T_{c}$ for  3\% reduction
of the ambient pressure lattice constant, as in the case of Ni$_{2}$MnSn.
Specifically, $T_{c}$ is increased by 40~K, a value smaller than but
still comparable to that for Ni$_{2}$MnSn and in reasonable agreement 
with experiment.
We find that any additional (i.e., in addition to that given by LSDA) 
Coulomb term $U$ on Pd-sites has a negligible effect on $T_c$ or its
pressure-dependence.   

The hockey-stick-like variation of the composition dependence of $T_c$ 
for Ni$_{1-x}$,Pd$_{x}$)$_{2}$MnSn alloys at 
ambient pressure is captured in the RPA, 
but not in the MFA. The RPA is equivalent to including the reaction field (or the Onsager cavity field) 
effect in the calculation of $T_c$
(see, for example, Cyrot\cite{cyrot} and references therein). Our results point to the importance of including
the reaction field effect in reproducing  the correct (experimentally observed) composition-dependence of $T_c$ in 
this alloy system.

The temperature can influence the above results, as discussed at the end of section~\ref{Form}. Increasing temperature leads to
an expansion of the lattice, compensating somewhat for  the pressure-caused contraction.
In essence, thus, the above reasoning should be applied to a net effective reduction of volume, 
considering both temperature and pressure effects. Lattice expansion effects and the effect of including the
Fermi-Dirac distribution function in performing the exchange integrals in Eq.~(2) have been discussed recently
by Alling \etal\cite{qsha2}. However, a complete $T\neq 0$ calculation, as described in section~\ref{Form}, 
has not yet been done. 

\subsubsection{(Ni$_{1-x}$,Pd$_{x}$)$_{2}$MnSn Heusler alloys}

\begin{figure}
\center \includegraphics[width=6.5cm]{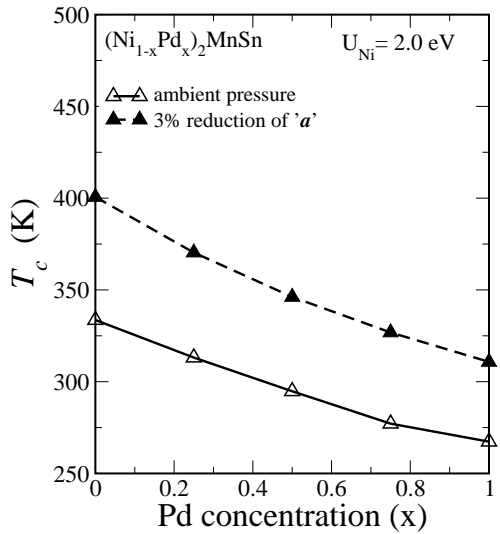}
\caption { The concentration-dependence of Curie temperatures (RPA)
for (Ni$_{1-x}$,Pd$_{x}$)$_{2}$MnSn alloys for ambient pressure and
for the pressure corresponding to the reduction of the alloy lattice
constant by 3\%. Model assumes electron correlations on Ni-sites
treated in the framework of the LSDA+U method.}
\label{f1}
\end{figure}

The results of the pressure-dependence of $T_{c}$ for (Ni,Pd)$_{2}$MnSn
alloys are shown in Fig.~\ref{f1}.
We have already mentioned that the ratio by which $T_{c}$ increases
with pressure for the end-point alloys is similar for both systems.
This is also reflected in the concentration-dependence of $T_{c}$, namely
a similar increase of $T_{c}$ under pressure over the whole concentration
range.
The monotonic and essentially linear concentration-dependence of $T_{c}$
under pressure reflects a similar trend of exchange interactions found for the
case of ambient pressure (see Ref.~\onlinecite{qha_th}). 
This similarity is due to the isoelectronic nature of Ni- and Pd-atoms coming from the
same column, but different rows ($3$- and $4$-$d$, respectively) of the Periodic Table.

\subsubsection{(Ni$_{1-x}$,Cu$_{x}$)$_{2}$MnSn Heusler alloys}

\begin{figure}
\center \includegraphics[width=6.5cm]{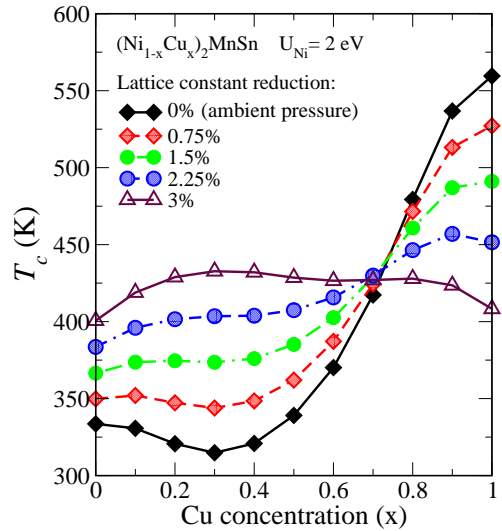}
\caption {(Color online)
The concentration-dependence of Curie temperatures (RPA)
for (Ni$_{1-x}$,Cu$_{x}$)$_{2}$MnSn alloys for ambient pressure and
for set of pressures corresponding to indicated reductions of the
alloy lattice constant. Model assumes electron correlations on Ni-sites
treated in the framework of the LSDA+U method.
}
\label{f2}
\end{figure}

Results of a similar study for (Ni,Cu)$_{2}$MnSn alloy over a set
of pressures realized by a linear reduction of the alloy lattice 
constant from the ambient value up to 3\% are shown in Fig.~\ref{f2}.
For comparison with experiments, results for the ambient pressure are slightly improved for the present 
LSDA+U model with respect to the LSDA results of our previous study
\cite{qha_th}. However, both LSDA and LSDA+U reveal the same general trend. 
Specifically, we observe two concentration regions, the first one 
for $x \le $0.4 in which  $T_{c}$ is essentially constant with a shallow 
minimum and that for $x > $0.4 where $T_{c}$ increases monotonically 
with $x$ .
With increasing pressure the behavior in the first concentration
region remains unchanged, but there is a gradual reduction of the
slope in the other concentration region (the Cu-rich end).
Such a development is clearly related to opposite trends in the
pressure-dependence of $T_{c}$ in Ni$_{2}$MnSn and Cu$_{2}$MnSn, a reduction of $T_{c}$
with pressure for Cu$_{2}$MnSn and an increase for Ni$_{2}$MnSn. 
The crossover from the positive derivative of the pressure-dependence
of $T_{c}$ to the negative one takes place around $x$=0.7.
Thus, for the pressure corresponding to a 3\% reduction of the
lattice constant we predict only a weak dependence of $T{_c}$ 
on the alloy composition.

\subsection{Qualitative understanding of results}

In this subsection we will present a qualitative understanding
of the composition-dependence of $T_c$ at ambient pressure as well as the effect of pressure on $T_{c}$ using the idea of
 Anderson's superexchange \cite{sasioglu,anderson} interaction and Stearns \cite{stearns} model of the
indirect (RKKY-type) interaction in ferromagnetic Heusler alloys.

The Mn-Mn exchange interaction, as in most other cases, can be divided into three groups: direct, indirect and superexchange.
Because the Mn-atoms are not the nearest neighbors of themselves and have a large separation ($d_{\rm{Mn-Mn}} >4$\AA), the 
direct interaction is not of any importance in Heusler alloys. The most important interaction turns out to be the
RKKY-type indirect interaction\cite{sasioglu}. Of somewhat minor importance is the 
superexchange interaction, introduced in a series of papers by Anderson\cite{anderson}.
Superexchange, which is also an indirect interaction, is antiferromagnetic, while the RKKY-type interaction can be
ferromagnetic or antiferromagnetic.
 Sasioglu \etal\cite{sasioglu} recently used these two interactions to interpret the results for some
Heusler alloys. Superexchange is related to possible transitions from occupied states 
at the Fermi level $E_F$ to unoccupied levels above, in the
 present case to the unoccupied minority Mn-bands.
Because the superexchange interaction is negative, the stronger the superexchange part, the smaller is the total exchange integral
(and $T_{c}$). 
The strength of the superexchange interaction depends on the density of
states (DOS) at $E_F$ and the energy separation between  $E_F$ 
and the unoccupied Mn-bands.
The larger the DOS at $E_F$ and/or the smaller the above energy separation, the
stronger is the superexchange.
This lends a qualitative understanding of the  increase of $T_{c}$ 
from Pd$_{2}$MnSn to Ni$_{2}$MnSn to Cu$_{2}$MnSn,  if the DOS mechanism 
dominates over the decreasing of energy separation in the above sequence of alloys 
(Fig.2 of Ref.~\onlinecite{qha_th}).
If we dope the Ni-sublattices with Cu-atoms in the  Ni$_{2}$MnSn host, the observed 
hockey-stick-like composition-dependence of $T_c$ can be understood as a competition of 
above two effects: for low Cu-content the decrease of $T_{c}$ with increasing Cu-concentration is
due to reduction of the above energy separation, causing an increase of the superexchange part
of the interaction.  For 
$x_{\rm Cu} > 0.35$ the DOS-effect starts to dominate, with decreasing DOS at $E_F$ and thus
smaller superexchange and larger total exchange interaction.   This leads
to higher  $T_{c}$ of Cu$_{2}$MnSn as compared to Ni$_{2}$MnSn.

In (Ni,Pd)$_{2}$MnSn alloy the DOS-effect is the only important effect over the entire concentration
range. Because the  Ni- and Pd-atoms have the same valency, the above mentioned energy separation does not change 
with concentration. There is a small decrease in the DOS at $E_F$ in Pd$_{2}$MnSn compared with Ni$_{2}$MnSn 
(Fig.2 of Ref.~\onlinecite{qha_th}),
decreasing the superexchange and increasing $T_{c}$.

As mentioned above, although the superexchange mechanism seems to support the observed results, it cannot be the driving
mechanism for the Heusler alloys under study. This is because of relatively large separation of the unoccupied minority Mn-peak    
above $E_F$ (the Mn-DOS just above $E_F$ is small). Of much-larger importance is the indirect interaction and its key feature
in ferromagnetic Heusler alloys was captured and successfully described by Stearns\cite{stearns} in a series of papers in
the late seventies. Stearns asserts that the magnetism of Heusler alloys (as well as bcc Fe\cite{stearns2}~) can be 
understood by dividing the
$d$-electrons into localized $d$-electrons ($d_l$) associated with narrow $d$-subbands and itinerant 
$d$-electrons ($d_i$) associated with (usually) one broader $d$-subband. The interaction between these localized 
and itinerant $d$-electrons is similar in nature to
RKKY interaction, originally formulated for the interaction
between localized moments in a free electron gas. The $d$-band gets narrower across (with increasing number of electrons)
a given series and broader with increasing row number (e.g. from $3d$ to $4d$).  Thus Stearns used her model to 
describe correctly the change of $T_c$ from
Ni$_2$MnSn to  Cu$_2$MnSn (an increase in $d_l$ causing a rise of $T_c$) and from Ni$_2$MnSn to  Pd$_2$MnSn
(a decrease of $T_c$ due to decreasing $d_l$). Uhl\cite{qha1} used Stearns model to explain the hockey-stick
appearance of the composition-dependence  of $T_c$ with increasing $x$ in (Ni$_{1-x}$,Cu$_{x}$)$_{2}$MnSn Heusler alloys.
With increasing Cu-concentration, the lattice dilates. Initially, this dilation causes a reduction in the strength of the
$d_i$-$d_l$ interaction, associated with increased Mn-Mn distance. However, the addition of Cu also causes an increase in the
number of $d_l$ and thus in the $d_i$-$d_l$ interaction. The resulting $T_c$ thus shows an initial decrease followed subsequently by
an increase (for further details, see Uhl~\cite{qha1}~). Thus both superexchange and indirect $d_i$-$d_l$ exchange interaction 
give rise to the same trend for the composition-dependence of $T_c$.

The pressure effect on $T_c$ can be understood in the light of the above explanation for the composition-dependence at
ambient pressure. Pressure reduces the lattice parameter, an effect opposite to the dilation caused by Cu-addition. Pressure also
broadens bands, reducing $d_l$ and increasing $d_i$, again opposite to what happens with increasing Cu-content. Thus on the Ni-rich
side increased pressure increases $T_c$ (opposite to the result with Cu-addition), and beyond a certain critical Cu-concentration
increasing pressure results in decreasing $T_c$ (opposite to the result with Cu-addition).

Similar reasoning explains the composition-dependence of ambient pressure $T_c$  and the pressure-dependence of $T_c$ in 
(Ni$_{1-x}$,Pd$_{x}$)$_{2}$MnSn alloys. Doping Ni-sublattice with Pd-atoms results in dilation of the lattice. In addition,
d-electrons in Pd are more delocalized than in Ni. Both of these will lead to a decrease of $T_c$ with increasing Pd concentration.
Pressure, via contraction of the lattice, reverses this effect. It appears that in this case 
the increasing strength of the interaction due to
reduction in the Mn-Mn distance supersedes the effect of the pressure-induced delocalization of the electrons. 

A couple of comments are in order. Stearns model is applicable to only ferromagnetic Heusler alloys. In several semi-Heusler alloys, or in general
in situations where the superexchange part of the interaction becomes important, Stearns model may fail to predict the results correctly.
It would not be incorrect  to point out that without  an accurate  first-principles 
study as presented here, it would be almost impossible to capture theoretically the details
of the composition- or pressure-dependence of $T_c$, although some qualitative 
understanding (especially, {\it a posteriori}) is possible.
In this sense the importance of first-principles studies like the present
one is irrefutable. 

\subsection{Spin-disorder induced resistivity}
\label{RR}

\begin{figure}
\center \includegraphics[width=6.5cm]{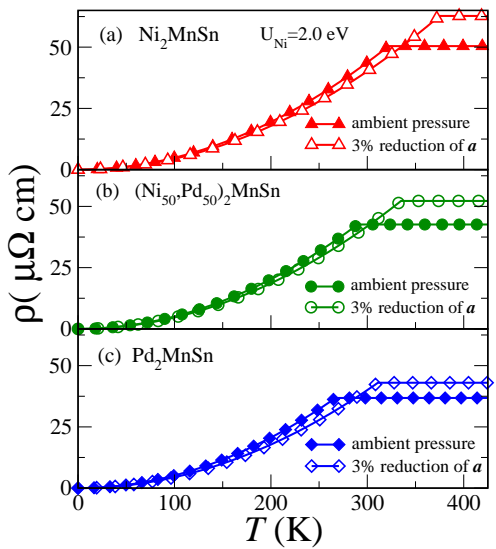}
\caption {(Color online) 
The temperature-dependence of the resistivity of the
(Ni$_{1-x}$,Pd$_{x}$)$_2$MnSn alloys for
ambient pressure (full symbols) and the pressure corresponding to the
reduction of the alloy lattice constant by 3\% (open symbols):
(a) $x=0$, (b) $x=0.5$, (c) $x=1$.
}
\label{f3}
\end{figure}

In magnetic alloys there are three contributions to the resistivity:
a temperature-independent residual resistivity due to the alloy 
disorder and other defects, and two temperature-dependent contributions, one due 
to electron-phonon and the other due to electron-magnon scatterings.
In the present case, the residual resistivity  is small because the
states corresponding to Ni-, Cu-, and Pd-atoms are well below the
alloy Fermi energy, \cite{qha_th} and thus influence states at the
Fermi energy (relevant for residual resistivity) only weakly.
The contribution due to phonons is well understood, and 
for Heusler alloys  around 
room temperature, \cite{resT} this contribution is known to be small.
The dominant contribution is thus due to spin-disorder. \cite{spindis1,spindis2}
The spin-disorder induced resistivity  is controlled by the spin-spin 
correlation function
which is small at and above  $T_{c}$,  so that a reasonable representation
of the situation is given by the DLM state, as explained in detail elsewhere.
\cite{qha_th}
Kasuya's $s$-$d$ interaction model\cite{spindis2} shows that the temperature-dependence of 
spin-disorder resistivity
is primarily quadratic and can be well-simulated by the $B T^{2}$-law.
In the present case we determine the constant $B$ from first principles
using $B=\rho(T_{c})/T_{c}^{2}$, where $\rho(T_{c})$ is identified with 
the resistivity of the DLM state.
The DLM state is described in the framework of the CPA as an
equiconcentration alloy with two types of  Mn atoms (A and B) with their moments pointing randomly up or down
(collinear disorder).  The corresponding resistivity can be 
properly calculated by the Kubo-Greenwood formula applied to this binary-Mn alloy.
The results for Ni$_{2}$MnSn, (Ni$_{0.5}$,Pd$_{0.5}$)MnSn, and Pd$_{2}$MnSn
alloys are shown in Fig.~\ref{f3}.  
It should be noted that the residual resistivity of the
(Ni$_{0.5}$,Pd$_{0.5}$)MnSn alloy in the ferromagnetic state is about two
orders of magnitude smaller than that in the DLM state. \cite{qha_th} 

The spin-disorder part of the resistivity is controlled by the
constant $B$.
For the cases shown in Fig.~\ref{f3}, $T_{c}$  increases under pressure.
Due to band-broadening under pressure and accompanying delocalization of the electrons,
$\rho(T)$ decreases in general for a given $T$ less than the ambient pressure $T_c$. However, $\rho(T_{c})$ 
(as well as $\rho(T)$ for $T\geq$ ambient pressure $T_c$) is higher under pressure because
of the increase in $T_c$.

To our knowledge, there are no available experimental results 
for the systems studied in this paper, although the results of
Austin and Mishra \cite{tc_pr3} for related Pd$_{2}$MnSb Heusler
alloy do indicate, as in this work as well, an increase of the resistivity 
under pressure for temperatures above $T_{c}$. 

Undoubtedly pressure has non-negligible  influence on the part of
the resistivity due to phonons.  This, however, is beyond the scope of the
present article.

\section{Conclusions}
\label{Con}

We have studied magnetic and  transport properties of quaternary 
Heusler alloys (Ni,T)$_{2}$MnSn (T=Cu, Pd) under pressure by means of  the
first-principles  density functional method.
In particular, we have investigated in detail the pressure-dependence
of the Curie temperature.
In agreement with experiments, we  obtain an increase of the Curie
temperature under pressure (reduction of the lattice constant)
for Ni$_{2}$MnSn and Pd$_{2}$MnSn.
On the other hand, Cu$_{2}$MnSn alloy exhibits a reduction of the Curie
temperature with applied pressure.
The results can be qualitatively understood as an interplay of two
effects: an increase of the bare exchange integrals with the volume reduction 
and the decrease of magnetic moments with pressure due to band broadening.

The concentration-dependence of the Curie temperature under pressure is
simple in (Ni,Pd)$_{2}$MnSn Heusler alloys where the increase of the
Curie temperature has an almost constant slope.
On the other hand, we predict a dramatic change of the Curie temperature
with pressure from the positive for Ni-rich (Ni,Pd)$_{2}$MnSn alloys to a
negative one in Cu-rich alloys, the crossover being around 70~\% of Cu. 
We emphasize that such a complex behavior cannot  be obtained
without an accurate calculation of the exchange interaction and a 
proper statistical treatment to compute $T_c$ (i.e., by going beyond the MFA).

A simple explanation of the behavior of the calculated $T_{c}$ under
pressure is given in the framework of Anderson's superexchange interaction and the
Stearns model of the indirect exchange interaction between itinerant and localized $d$-electrons.

We have also investigated the pressure-dependence of the spin-disorder
related part of the resistivity, which dominates at higher temperatures,
 the residual resistivity due to alloy disorder being small and important
only for very low temperatures.
We have found that the resistivity at the Curie temperature increases
with pressure, but this result is due to the increase of the
Curie temperature with pressure itself. For temperatures below the ambient pressure
$T_{c}$ the resistivity decreases under pressure, as would be expected due to
band broadening.

\begin{acknowledgments}
This work was supported by a grant from the Natural Sciences and 
Engineering Research Council of Canada.
J.K. and V.D. acknowledge financial support from AV0Z 10100520 and 
the Czech Science Foundation (202/09/0775).
The work of I.T. has been supported by the Ministry of Education
of the Czech Republic (Grant No. MSM 0021620834) and by the Czech 
Science Foundation (Grant No. 202/09/0030).

\end{acknowledgments}


\end{document}